\documentstyle[twoside,fleqn,espcrc2]{article}

\newcommand{\AmS}{{\protect\the\textfont2
  A\kern-.1667em\lower.5ex\hbox{M}\kern-.125emS}}

\title{Non-analyticity of the Callan-Symanzik  $\beta$-function of O($N$) models.}
\author{Andrea Pelissetto${}^{\rm a}$ and 
Ettore Vicari\address{Dipartimento di Fisica dell'Universit\`a di Pisa, Italy.}
}
       
\begin{document}

\begin{abstract}
In the framework of the $1/N$ expansion we show that the Callan-Symanzik 
$\beta$-function associated with the four-point coupling $g$
is non-analytic at its zero, i.e. at the fixed-point value $g^*$ 
of $g$. This singular behavior can be interpreted  by 
renormalization group arguments, and written in terms of scaling
correction exponents.

We obtain accurate determinations of $g^*$ in 3-$d$ and 2-$d$ 
by exploiting two alternative approaches:
the $\epsilon$-expansion in the $\phi^4$ formulation of the O($N$) model, and 
the high-temperature expansion of the lattice $N$-vector 
(O($N$) nonlinear $\sigma$) model.
These results are compared with the available estimates by other approaches,
such as the fixed-dimension perturbative expansion, Monte Carlo simulations,
etc...

We also present results for the $n$-point renormalized coupling constants that 
parameterize the behavior of the effective potential in the 
high- and low-temperature phases.

\end{abstract}

\maketitle

The renormalization-group theory of critical phenomena
provides a description of statistical models 
in the neighbourhood of the critical point.
For ${\rm O}(N)$ models calculations are based on
the $\phi^4$-field theory.
A strategy, which has been largely employed in the
study of the symmetric phase, relies on a perturbative expansion
in powers of the zero-momentum four-point coupling $g$
performed at fixed dimension.
The theory is renormalized at zero momentum
by requiring 
\begin{eqnarray}
&&\Gamma^{(2)}(p)_{\alpha\beta} = \delta_{\alpha\beta}\;
Z_G^{-1} \left[ m^2+p^2+O(p^4)\right]
\label{s2e7a}  \\
&&\Gamma^{(4)}(0,0,0,0)_{\alpha\beta\gamma\delta} =
Z_G^{-2} \,m\, g \,\delta_{\alpha\beta\gamma\delta}/3
\label{s2e7b}
\end{eqnarray}
where
$\delta_{\alpha\beta\gamma\delta}
=\delta_{\alpha\beta}\delta_{\gamma\delta}
+\delta_{\alpha\gamma}\delta_{\beta\delta}
+\delta_{\alpha\delta}\delta_{\beta\gamma}$.
When $m\rightarrow 0$ the coupling $g$ is driven toward
an infrared stable zero $g^*$ of the Callan-Symanzik $\beta$-function
\begin{equation}
\beta(g)\equiv m{\partial g/ \partial m}|_{g_0,\Lambda}.
\end{equation}
$g^*$ is also obtained as the critical limit of 
\begin{equation}
g_\sigma =-{3N\over N+2} {\chi_4 \over \chi^2\xi^d }\longrightarrow g^*
\label{deff2}
\end{equation}
where $\chi$ is the magnetic susceptibility, $\xi$ the second-moment
correlation length, and 
$\chi_4 =  \sum_{i,j,k}
\langle \phi(0)\cdot \phi(x_i) \phi(x_j)\cdot \phi(x_k) \rangle_c.$
We recall that $N$-vector (nonlinear O($N$) $\sigma$) and $\phi^4$ models 
describe the same critical behavior.

In the framework of the $1/N$ expansion 
the analysis of the next-to-leading order shows 
 that the Callan-Symanzik 
$\beta$-function is non-analytic at its zero, i.e. at the fixed-point value $g^*$ 
of $g$~\cite{gre}. 
The large-$N$ result agrees with the singular behavior
\begin{eqnarray}
&&\beta(g) = -\omega  (g^{*} - g) + \;\; \hbox{\rm analytic terms} 
\nonumber \\ 
&&   +  c_1 (g^{*} - g)^{1+{1\over \Delta}} + \ldots 
  +  d_1 (g^{*} - g)^{\Delta_2\over \Delta} + \ldots 
\label{rg}
\end{eqnarray}
($\Delta=\omega\nu$ and $\Delta_2$ are scaling correction 
exponents) that can be derived using 
renormalization group arguments~\cite{Nickel}. 

\begin{table*}[hbt]
\setlength{\tabcolsep}{1.15pc}
\caption{
Three-dimensional  estimates of $\bar{g}^*\equiv g^* (N+8)/(48\pi)$.
For the Ising model we also mention the recent Monte Carlo estimate~\cite{B-F-M-M}
$\bar{g}^*=1.397(14)$ (actually derived by us using the data kindly made available by the
authors).
A more complete list of the available estimates of $g^*$ 
can be found in Ref.~\cite{gre}.
}
\label{t1}
\begin{tabular*}{\textwidth}{lllllll}
\hline
                   \multicolumn{1}{c}{$N$} 
                 & \multicolumn{1}{c}{$\epsilon$-exp.\cite{gre}} 
                 & \multicolumn{1}{c}{$g$-exp.\cite{G-Z}} 
                 & \multicolumn{1}{c}{$g$-exp.\cite{M-N}} 
                 & \multicolumn{1}{c}{H.T.\cite{gre}} 
                 & \multicolumn{1}{c}{H.T. cubic\cite{B-C-g3}} 
                 & \multicolumn{1}{c}{H.T. bcc\cite{B-C-g3}} \\
\hline
0 & $1.390(17)$ & $1.413(6) $ & $1.39 $ & $1.393(20)$& $1.388(5) $ & $1.387(5) $\\
1 & $1.397(8) $ & $1.411(4) $ & $1.40 $ & $1.406(9) $& $1.408(7) $ & $1.407(6) $\\
2 & $1.413(13)$ & $1.403(3) $ & $1.40 $ & $1.415(11)$& $1.411(8) $ & $1.406(8) $\\
3 & $1.387(7) $ & $1.391(4) $ & $1.39 $ & $1.411(12)$& $1.409(10)$ & $1.406(8) $\\
4 & $1.366(15)$ & $1.377(5) $ &         & $1.396(16)$& $1.392(10)$ & $1.394(10)$ \\ 
\hline
\end{tabular*}
\end{table*}

\begin{table*}[hbt]
\setlength{\tabcolsep}{1.35pc}
\caption{Two-dimensional estimates of $\bar{g}^*\equiv g^* (N+8)/(24\pi)$.
We recall that O($N$) $\sigma$ models with $N\geq 3$ are asymptotically free,
thus $\beta_c=\infty$. 
}
\label{t2}
\begin{tabular*}{\textwidth}{llllll}
\hline
                   \multicolumn{1}{c}{$N$} 
                 & \multicolumn{1}{c}{$\epsilon$-exp.\cite{gre}} 
                 & \multicolumn{1}{c}{$g$-exp.} 
                 & \multicolumn{1}{c}{H.T.} 
                 & \multicolumn{1}{c}{M.C.} 
                 & \multicolumn{1}{c}{$1/N$-exp.} \\
\hline
0 & 1.69(7) &                        & 1.679(3)~\cite{B-C-g2}  &                      &  \\
1 & 1.75(5) & 1.85(10)~\cite{L-Z}    & 1.7540(2)~\cite{B-C-g2} & 1.71(12)~\cite{K-P} &  \\
2 & 1.79(3) &                        & 1.810(10)~\cite{B-C-g2} & 1.76(3)~\cite{Kim}   &  \\
3 & 1.72(2) & 1.749(16)~\cite{F-etal}& 1.724(9)~\cite{gre}      & 1.73(3)~\cite{Kim}  & 1.758 \\
4 & 1.64(2) &                        & 1.655(16)~\cite{B-C-g2} &                      & 1.698 \\ 
\hline
\end{tabular*}
\end{table*}

A precise determination of $g^*$ is crucial in the 
field-theoretic approach
based on the $g$-expansion, where the critical exponents 
are obtained by evaluating appropriate (resummed)
anomalous dimensions at $g^*$.
In this approach the resummation of the $g$-expansion  
is usually performed following the 
 Le Guillou Zinn-Justin (LZ) procedure~\cite{L-Z}, 
which  assumes the analyticity of the $\beta$-function.
The presence of confluent singularities may then  cause a slow convergence to
the correct fixed-point value, leading to an underestimation of the uncertainty. 
A more general analysis explicitly allowing for the presence of
confluent singularities would slightly change the value of $g^*$ 
for small values of $N$ and consequently
the values of the critical exponents~\cite{M-N}.
It is therefore important
to exploit other approaches to the study of O($N$) models,
which can provide a check of the estimates of $g^*$ 
from the resummations of the $g$-expansion.
We considered two alternative approaches:
the $\epsilon$-expansion in the continuum $\phi^4$ formulation
and the high-temperature (HT) expansion of the $N$-vector (lattice $O(N)$ 
$\sigma$) model.
We extended the $\epsilon$-expansion of $g^*$ to $O(\epsilon^4)$.
Accurate estimates of $g^*$ in 3-$d$ 
and 2-$d$ were obtained by a constrained analysis of the $\epsilon$-series 
using its known values at lower dimensions.
Moreover we reanalyzed the available
HT expansion of $g_\sigma$ in the  $N$-vector models, by a method
able to handle the leading confluent singularity (for a more recent analysis using 
longer series see Ref.~\cite{B-C-g3}).
In Table~\ref{t1} and \ref{t2} we present our 3-$d$ and 2-$d$ results
respectively. For comparison
we also report some of the available estimates from other approaches.
The agreement among the various estimates of $g^*$ is globally good.

The results in Table~\ref{t1} indicate that the systematic error 
in the LZ resummation  
due to the non-analytic terms in Eq.~(\ref{rg})  
should be small. This may be explained by the 
fact that, for small values of $N$, 
the exponents in Eq.~(\ref{rg}) are close 
to integer numbers, indeed
$\Delta_2/\Delta\simeq 2$, $\Delta_3/\Delta\simeq 3$,
and $1+1/\Delta\simeq 3$. 
However, the results for $N=0,1$ are slightly lower than the estimates given by the
LZ resummation of the $O(g^7)$ series of $\beta(g)$,
thus favouring the more general analysis of Ref.~\cite{M-N}.
This would lead to a small change in the estimates of the critical exponents.
For instance, in the case $N=0$ (self avoiding walks)
the resummation of the $O(g^7)$ series of $\gamma(g)$
evaluated at $\bar{g}^*=1.413(6)$ gives $\gamma(\bar{g}^*)\simeq 1.160$.
A lower value $\bar{g}^*\simeq 1.39$, as indicated by our calculations,
would lead to $\gamma(\bar{g}^*=1.39)\simeq 1.158$,
in substantial agreement with the recent result
of Monte Carlo simulations $\gamma=1.1575(6)$~\cite{C-C-P}
and with the analysis of the
$\epsilon$-expansion: $\gamma\simeq 1.158$.

\begin{table*}[hbt]
\setlength{\tabcolsep}{0.5pc}
\caption{
Three-dimensional estimates of $r_6$ and $r_8$.
A more complete list of the estimates of $r_{2j}$
is reported in \cite{effpot}.
}
\label{t3}
\begin{tabular*}{\textwidth}{ccr@{}lr@{}lr@{}lr@{}lcr@{}lr@{}lr@{}lr@{}l}
\hline
\multicolumn{1}{c}{$N$}&
\multicolumn{1}{c}{}&
\multicolumn{8}{c}{$r_6$}&
\multicolumn{1}{c}{}&
\multicolumn{8}{c}{$r_8$}\\
\multicolumn{1}{c}{}&
\multicolumn{1}{c}{}&
\multicolumn{2}{c}{$\epsilon$-exp.}&
\multicolumn{2}{c}{$g$-exp.}&
\multicolumn{2}{c}{ERG}&
\multicolumn{2}{c}{H.T.}&
\multicolumn{1}{c}{}&
\multicolumn{2}{c}{$\epsilon$-exp.}&
\multicolumn{2}{c}{$g$-exp.}&
\multicolumn{2}{c}{ERG}&
\multicolumn{2}{c}{H.T.}\\
\hline
1 & $\quad$ & 2&.058(11)& 2&.053(8) & 2&.064(36) & 1&.99(6) &  
$\qquad$ &  2&.48(28) & 2&.47(25) & 2&.47(5) & 2&.7(4) \\

2 & $\quad$ & 1&.94(11) & 1&.967 & 1&.83  & 2&.2(6)\ &  
$\qquad$ &  3&.5(1.3) & & & 1&.4 & &  \\

3 & $\quad$ & 1&.84(9)  & 1&.880 & 1&.74 & 2&.1(6) &  
$\qquad$ &  2&.1(1.0) & & & 0&.84 & &    \\ 

4 & $\quad$ & 1&.75(7)  & 1&.803 & 1&.65 & 1&.9(6) &  
$\qquad$ &  1&.2(1.0) & & & 0&.33 & &    \\ 
\hline
\end{tabular*}
\end{table*}

The effective potential is widely used 
in the field-theoretic description of fundamental interactions 
and phase transitions.
In statistical physics it represents the 
free-energy density ${\cal F}$ as a function of the order parameter.
${\cal F}$ can be expanded  
in powers of the renormalized magnetization $\varphi$
\begin{equation}
\Delta {\cal F} =
\sum_{j=1} m^{2j+(1-j)d}
{1\over (2j)!} g_{2j} \varphi^{2j}
\label{freeeng}
\end{equation}
where $\Delta {\cal F}= 
 {\cal F}(\varphi) -  {\cal F}(0)$, and
$g_{2j}$ are the zero-momentum $2j$-point renormalized coupling.
By definition $g_2=1$ and $g_4=g$.
Setting $\varphi = m^{(d-2)/2} z / \sqrt{g}$
and $r_{2j} = {g_{2j}/g^{j-1}}$ we write
\begin{equation}
\Delta {\cal F} = {m^d\over g}
\left( {1\over 2} z^2 + {1\over 4!} z^4 + 
\sum_{j=3} {1\over (2j)!} r_{2j} z^{2j}\right)
\end{equation}

In order to evaluate the first few $r_{2j}$ 
we performed a constrained analysis of the $\epsilon$-expansion
of $r_{2j}$. In Table~\ref{t3} we report our 3-$d$
results for $r_6$ and $r_8$. We compare them with
some of the available estimates from other approaches,
such as $d=3$ $g$-expansion~\cite{G-Z,S-O-U-K}, approximate solution
of the exact renormalization group equation (ERG)~\cite{Morris,T-W},
high-temperature expansion~\cite{B-C-gj,Reisz}.
In the case of the Ising model also $r_{10}$ has been roughly estimated:
$r_{10}= -20(15)$.

In two dimensions an analysis of the HT expansion of the free-energy 
of the Ising model in the presence of an external field gave
$r_6=3.678(2)$, $r_8 = 26.0(2)$ and $r_{10}=275(15)$. These numbers
compare well with the estimates $r_6=3.69(4)$ and $r_8 = 26.4(1.0)$ obtained 
from our constrained analysis of the $\epsilon$-expansion.
Moreover we obtained 
$r_6=3.54(7)$ and $r_8 = 25.1(2.0)$ for $N=2$ (2-$d$ XY model),
and $r_6=3.33(6)$ and $r_8 = 20.3(1.7)$ for $N=3$.

In the broken phase of the 3-$d$ Ising model 
the effective potential at the coexistence curve can be expanded as
\begin{equation}
{\cal F}(\varphi) - 
 {\cal F}(\varphi_0) =
\sum_{j=2} m^{d-j(d-2)/2}
{1\over j!} g _{j}^- (\varphi-\varphi_0)^j
\label{frbr}
\end{equation}
where $g_{j}^-$ are the zero-momentum $j$-point renormalized coupling
in the broken phase ($g_2^-=1$ by definition).
A constrained analysis of the $\epsilon$-expansion~\cite{brphase} gave 
the estimates $g_3^-=13.06(12)$ and $g_4^-=75(7)$.
The parametrization (\ref{frbr}) does not apply to the case
$N\neq 1$, due to the presence of Goldstone bosons. 
One indeed finds 
\begin{equation}
{\cal F}(\varphi) - {\cal F}(\varphi_0)
\approx c\left( \varphi^2-\varphi_0^2\right)^{d/(d-2)}
\end{equation}
In 3-$d$, where $d/(d-2)=3$,
logarithms appear in the corrections to the 
leading behavior.


\begin{thebibliography}{9}

\bibitem{gre} A. Pelissetto, E. Vicari,
Nucl. Phys. {\bf B519} (1998) 626.

\bibitem{Nickel}
B.~G.~Nickel, in {\em Phase Transitions},
M.~L\'evy, J.~C.~Le~Guillou and J.~Zinn-Justin eds.,
(Plenum, New York and London, 1982).

\bibitem{L-Z} J.~C.~Le Guillou, J.~Zinn-Justin,
Phys.\ Rev.\ {\bf B21} (1980) 3976.

\bibitem{M-N}
D.B. Murray,  B.G.~Nickel, Guelph University report (1991).

\bibitem{B-C-g3} P. Butera, M. Comi, hep-lat/9805025.

\bibitem{G-Z}
R.~Guida, J.~Zinn-Justin, cond-mat/9803240.

\bibitem{B-F-M-M}
H.G. Ballesteros, et al,
hep-lat/9805022; private communications.

\bibitem{F-etal}
M.~Falcioni, et al,
Nucl.\ Phys.\ {\bf B225} (1983) 313.

\bibitem{B-C-g2}
P.~Butera, N.~Comi, Phys. Rev. {\bf B54}  (1996) 15828. 

\bibitem{K-P}
J.~K.~Kim, A.~Patrascioiu, Phys.  Rev.  {\bf D47}  (1993) 2588. 

\bibitem{Kim}
J.~ Kim, Phys. Lett. {\bf B345} (1995) 469.

\bibitem{C-C-P}
S.~Caracciolo, M.~S.~Causo, A.~Pelissetto, 
Phys. Rev. {\bf E57} (1998) R1215.

\bibitem{effpot} A. Pelissetto,  E. Vicari,
Nucl. Phys. {\bf B522} (1998) 605.

\bibitem{S-O-U-K} A.~I.~Sokolov, et al,
hep-th/9808011.

\bibitem{Morris}
T.~Morris, Nucl. Phys. {\bf B495} (1997) 477.

\bibitem{T-W}
N.~Tetradis, C.~Wetterich, Nucl. Phys. {\bf B422} (1994) 541.

\bibitem{B-C-gj}
P.~Butera, N.~Comi, Phys.\ Rev.\ {\bf E55}  (1997) 6391.

\bibitem{Reisz}
T.~Reisz, Phys. Lett.  {\bf B360} (1995) 77.

\bibitem{brphase} A. Pelissetto, E. Vicari,
cond-mat/9805317.




\end{thebibliography}
\end{document}